\documentclass[pra,aps,showpacs,twocolumn,superscriptaddress]{revtex4}
\usepackage{amsfonts}
\usepackage{amsmath}
\usepackage{amssymb}
\usepackage{graphicx}
\usepackage{txfonts}

\begin{document}

\title{Quantum Fisher information as signature of superradiant quantum phase
transition}
\author{T. L. Wang}
\affiliation{Department of Physics, Beijing Jiaotong University, Beijing 100044, China}
\date{\today }
\author{L. N. Wu}
\thanks{Present address: Department of Physics, Tsinghua University, Beijing
100084, China.}
\affiliation{Department of Physics, Beijing Jiaotong University, Beijing 100044, China}
\date{\today }
\author{W. Yang}
\affiliation{Beijing Computational Science Research Center, Beijing 100084, China}
\author{G. R. Jin}
\email{grjin@bjtu.edu.cn}
\affiliation{Department of Physics, Beijing Jiaotong University, Beijing 100044, China}
\author{N. Lambert}
\affiliation{Advanced Science Institute, RIKEN, Wako-shi, Saitama 351-0198, Japan}
\author{F. Nori}
\affiliation{Advanced Science Institute, RIKEN, Wako-shi, Saitama 351-0198, Japan}
\affiliation{Physics Department, University of Michigan, Ann Arbor, MI 48109-1040, USA}
\date{\today }

\begin{abstract}
The single-mode Dicke model is well-known to undergo a quantum phase
transition from the so-called normal phase to the superradiant phase
(hereinafter called the ``superradiant quantum phase transition"). Normally,
quantum phase transitions are closely related to the critical behavior of
quantities such as entanglement, quantum fluctuations, and fidelity. In this
paper, we study the quantum Fisher information (QFI) of the field mode and that
of the atoms in the ground state of the Dicke Hamiltonian. For finite and
large enough number of atoms, our numerical results show that near the
critical atom-field coupling, the QFIs of the atomic and the field
subsystems can surpass the classical limits, due to the appearance of
nonclassical squeezed states. As the coupling increases far beyond the
critical point, the two subsystems are in highly mixed states, which degrade
the QFI and hence the ultimate phase sensitivity. In the thermodynamic
limit, we present analytical results of the QFIs and their relationships
with the reduced variances. For each subsystem, we find that there is a
singularity in the derivative of the QFI at the critical point, a clear
signature of quantum criticality in the Dicke model.
\end{abstract}

\maketitle

\section{Introduction}

Quantum phase transitions in many-body systems are of fundamental interest~\cite{Sachdev} and have potential applications in quantum information~\cite{Osterloh,GVidal,JVidal,Tsomokos,Shi,Lambert} and quantum metrology~\cite{Katori,Giovannetti,Zanardi,Invernizzi,Ma&Wang,Gammelmark,Genoni12,Vitali}.
Consider, for instance, a collection of $N$ two-level atoms interacting with a single-mode bosonic field, described by the Dicke model (with $\hbar =1$ throughout this paper)~\cite{Dicke}:
\begin{equation}
\hat{H}=\omega \hat{b}^{\dagger }\hat{b}+\omega _{0}\hat{J}_{z}+\frac{%
\lambda }{\sqrt{N}}(\hat{b}^{\dagger }+\hat{b})(\hat{J}_{+}+\hat{J}_{-}),
\label{DHam1}
\end{equation}%
where $\hat{b}$ and $\hat{b}^{\dagger}$ are annihilation and creation
operators of the bosonic field with oscillation frequency $\omega$, which
is nearly resonant with the atomic energy splitting $\omega _{0}$. The
collective spin operators $\hat{J}_{\pm }\equiv \hat{J}_{x}\pm i\hat{J}%
_{y}=\sum_{k}\hat{\sigma}_{k}^{\pm }$ and $\hat{J}_{z}=\sum_{k}\hat{\sigma}%
_{k}^{z}/2$ obey the SU(2) Lie algebra, where $\hat{\sigma}_{k}^{\pm }$ and $%
\hat{\sigma}_{k}^{z}$ are Pauli operators of the $k$-th atom. The atom-field
coupling strength $\lambda \varpropto \sqrt{N/V}$ depends on the atomic
density $N/V$. For a finite number of atoms $N$ ($\equiv 2j$), the
Hamiltonian~(\ref{DHam1}) commutes with the parity operator $\hat{\Pi}=\exp
[i\pi(\hat{b}^{\dagger}\hat{b}+\hat{J}_{z}+j)]$, due to $\hat{\Pi}^{\dag}%
\hat{J}_{x}\hat{\Pi}=-\hat{J}_{x}$ and $\hat{\Pi}^{\dag}\hat{b}\hat{%
\Pi}=-\hat{b}$~\cite{Baumann}. As a result, the ground state of the finite-$N$
Dicke model $|g\rangle $ does not exhibit any singularity and degeneracy.
This can be understood by expanding $|g\rangle $ in the basis $\{|n\rangle
|j,m\rangle\}$~\cite{EB03,Qing-Hu}, where $|n\rangle $ and $|j,m\rangle $
(with $m\in \lbrack -j$, $+j]$) are the Fock states and the eigenvectors of $%
\hat{J}_{z}$, respectively. For vanishing atom-field coupling strength $%
\lambda $, the ground state $|g\rangle =|0\rangle |j,-j\rangle $ has a
positive parity $\langle \hat{\Pi}\rangle =+1$; Similarly for $\lambda >0$,
due to the conserved parity, the ground state $|g\rangle $ consists of states with \emph{even} number $n+m+j$~\cite{EB03,Qing-Hu}, which results in
vanishing coherence (i.e., $\langle\hat{J}_{x}\rangle=\langle\hat{b}\rangle=0$). However, in the thermodynamic limit (for finite $N/V$ as $%
N $, $V\rightarrow\infty$), the parity symmetry is spontaneously broken
and the ground states with parities $\pm 1$ become degenerate in the
superradiant phase (i.e., the symmetry-broken phase at $\lambda \geq \lambda
_{\mathrm{cr}}=\sqrt{\omega _{0}\omega }/2$)~\cite{Hillery,No-go,EB03,Qing-Hu,YLi,Chen,Dimer,Song,Huang,Esslinger,Nagy,Ciuti,Marquardt}, leading to bifurcation of $\langle\hat{J}_{x}\rangle$ and that of $\langle\hat{b}\rangle$~\cite{Baumann}.

Unlike the traditional phase transition of the Dicke model at a finite
temperature~\cite{HL&WH}, the superradiant quantum phase transition is
driven by quantum fluctuations in the large-$N$ limit. It is natural to ask in what different ways one can characterize such a quantum phase transition in a realistic system. Several quantities, with various degrees of
experimental accessibility, have been shown to be sensitive to the quantum
phase transitions, such as the von Neumann entropy~\cite{Lambert}, the
fidelity~\cite{Fidelity}, and more recently the quantum fluctuations of the
field~\cite{Hur}.

In this paper, we investigate the quantum Fisher information (QFI) of the
field state $\hat{\rho}_{B}={\mathrm{Tr}}_{A}(|g\rangle \langle g|)$ and
that of the atomic state $\hat{\rho}_{A}={\mathrm{Tr}}_{B}(|g\rangle \langle
g|)$, where $\mathrm{Tr}_{A}$ ($\mathrm{Tr}_{B}$) is the partial trace of the
ground state $|g\rangle $ over the atomic (bosonic field) degrees of freedom. In
quantum metrology, the QFI is one of central quantities to qualify the input
state~\cite{Helstrom,Braunstein94}, especially in March-Zehnder
(or, equivalently, Ramsey) interferometer-based phase or parameter
estimation. The achievable phase sensitivity is well-known to be limited by
the quantum Cram\'{e}r-Rao bound $\delta\varphi_{\min}\propto 1/\sqrt{F}$, where the QFI $F$ depends on the input state and the phase-shift generator~\cite{Braunstein94,Smerzi09,Genoni}. Here, we show that near the
critical point $\lambda_{\mathrm{cr}}$, the QFI of $\hat{\rho}_{A,B}$ for the finite-%
$N$ Dicke model can surpass the classical limit due to the nonclassical
squeezed properties of the ground state. As the coupling strength $\lambda
\gg \lambda _{\mathrm{cr}}$, both $\hat{\rho}_{A}$ and $\hat{\rho}_{B}$ become highly
mixed states, which leads to the QFI of the field returning to the classical
limit, while for the atoms the QFI tends to be zero. In the thermodynamic
limit, we discover that there exists analytical relationships between the
QFIs and the reduced variances, which show clearly the squeezing-induced
enhancement of the QFIs. More interestingly, we find that the derivative of $%
F$ for each subsystem is divergent at $\lambda =\lambda _{\mathrm{cr}}$, similar to
the fidelity of the ground state $|g\rangle$ in the one dimensional Ising
chain~\cite{Fidelity}. This finding suggests that the QFI could be useful as
a sensitive probe of quantum phase transitions~\cite{Ma&Wang}.


\section{Quantum Fisher information in finite-$N$ Dicke model}

We first examine the field state $\hat{\rho}_{B}={\mathrm{Tr}}%
_{A}(|g\rangle\langle g|)$ of the finite-$N$ Dicke model by numerically
evaluating the QFI with respect to $\hat{\rho}_{B}(\varphi)=e^{i\varphi\hat{G}}\hat{\rho}_{B}e^{-i\varphi\hat{G}}$, where $\varphi$ is an unknown phase shift and $\hat{G}$ is the phase-shift generator ($=\hat{b}^{\dagger}\hat{b}$
for the single-mode field~\cite{Genoni}). In general, the field state $\hat{\rho}_{B}(\varphi)$ is a mixed state and the QFI is given by ~\cite{Braunstein94,Smerzi09,Genoni,Ma11,Knysh,Zhang13}
\begin{equation}
F=4\sum_{n}p_{n}(\Delta \hat{G})_{n}^{2}-\sum_{m\neq n}\frac{8p_{m}p_{n}}{
p_{m}+p_{n}}|\langle \psi _{m}|\hat{G}|\psi _{n}\rangle |^{2},  \label{QFI}
\end{equation}
where the weights $\{p_{n}\}$ are \textit{nonzero} eigenvalues of $\hat{\rho}%
_{B}$, and $\{|\psi_{n}\rangle\}$ are the corresponding eigenvectors. The
first term of Eq.~(\ref{QFI}) is a weighted average over the QFI for each
pure state $|\psi_{n}\rangle$, and the variance $(\Delta \hat{G}
)_{n}^{2}\equiv\langle\psi_{n}|\hat{G}^{2}|\psi_{n}\rangle-|\langle\psi_{n}|\hat{G}|\psi_{n}\rangle|^{2}$. The second term is simply a negative
correction (c.f. Ref.~\cite{Zhang13}). For a pure coherent state $|\alpha
\rangle$, with mean number of bosons $\bar{n}=|\alpha|^{2}$, we obtain the
QFI of the bosonic field, denoted by $F_{B}$ hereafter, $F_{B}=4(\Delta \hat{%
b}^{\dagger}\hat{b})^{2}=4\bar{n}$ and hence the ultimate sensitivity $%
\delta\varphi_{\min}^{\mathrm{cl}}=1/(2\sqrt{\bar{n}})$, which corresponds
to the classical (or shot-noise) limit. A sub shot-noise-limited phase sensitivity with $\delta \varphi <\delta \varphi_{\min}^{\mathrm{cl}}$ is achievable provided that $F_{B}>4\bar{n}$, which has been shown a nonclassical criterion of $\hat{\rho}_B$ for the single-mode linear interferometer~\cite{Rivas}.

The atoms in the ground state $\hat{\rho}_{A}={\mathrm{Tr}}_{B}(|g\rangle\langle g|)$ can also be used as a probe in a standard Ramsey
interferometer. Since the orientation of the atomic spin $\langle \mathbf{\hat{J}}\rangle$ is along the $\hat{J}_{z}$ axis (due to $\langle \hat{J}%
_{+}\rangle =0$), to precisely estimate the atomic transition frequency $\omega _{0}$, a $\pi /2$ pulse is required to rotate the atomic spin about
the $\hat{J}_{y}$ axis. After a free evolution $\tau$, the phase shift $\varphi=\omega_{0}\tau$ is accumulated, leading to the atomic state $\hat{%
\rho}_{A}(\varphi )=e^{i\varphi \hat{J}_{z}}e^{i\pi\hat{J}_{y}/2}\hat{\rho}_{A}e^{-i\pi\hat{J}_{y}/2}e^{-i\varphi \hat{J}_{z}}$, where $e^{i\pi\hat{J}_{y}/2}$ and $e^{i\varphi\hat{J}_{z}}$ represent the action of the pulse
and the phase accumulation, respectively. Again, the QFI of the reduced
atomic state $\hat{ \rho}_{A}(\varphi)$ is given by Eq.~(\ref{QFI}), where
the phase-shift generator $\hat{G}$ is replaced by $\hat{J}_{x}$ and \{$|\psi _{n}\rangle$\} are eigenvectors of $\hat{\rho}_{A}$ with nonzero
weights $p_{n}$. For a coherent spin state $|j,-j\rangle=|\downarrow\rangle^{\otimes N}$, we have the QFI of the atoms $F_{A}=4(\Delta\hat{J}_{x})^{2}=N$ so the sensitivity is limited by $\delta\varphi_{\min}^{\mathrm{cl}}=1/\sqrt{N}$ (i.e., the classical limit). Hereafter, we denote $F_{A}$ as the QFI of the atoms, to distinguish it from that of the bosonic field $F_{B}$.

\begin{figure}[hbpt]
\centerline{
\includegraphics[width=0.95\columnwidth,angle=0]{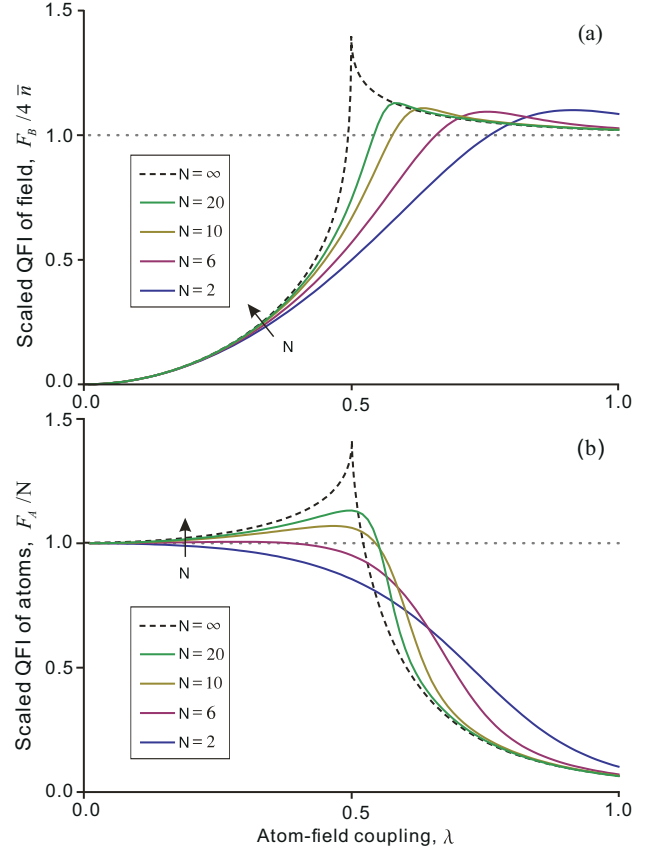}}
\caption{(color online) Scaled quantum Fisher information of the bosonic
field $F_{B}/(4\bar{n})$ (a) and that of the atoms $F_{A}/N$ (b) as a
function of the coupling strength $\protect\lambda $ for a finite number of
atoms $N=2$, $6$, $10$, and $20$, as indicated by the arrow. Horizontal
dotted lines: the classical (or shot-noise) limit for the field mode $F_{B}=4 \bar{n}$ (with mean number of bosons $\bar{n}$) and that of the
atoms $F_{A}=N$. Dashed lines: analytical results of the QFIs in the
thermodynamic limit (i.e., $N=\infty$). For each state $\hat{\rho}_{A,B}$, the derivative of the QFI has a singularity at the critical point $\lambda_{\mathrm{cr}}$. Other parameters: the critical coupling $\lambda _{\mathrm{cr}}\equiv \sqrt{\omega\omega_{0}}/2=1/2$
for resonant condition $\protect\omega=\omega_{0}=1$.}
\label{FIG1}
\end{figure}

In Fig.~\ref{FIG1}, we plot the scaled QFI of the field $F_{B}/(4\bar{n})$
and that of the atoms $F_{A}/N$ as a function of the atom-field coupling
strength $\lambda$. The QFI of the field vanishes at the atom-field coupling
$\lambda=0$, since the bosonic field is in vacuum (i.e., $\hat{\rho}_{B}$ $%
=|0\rangle\langle 0|$). By contrast, the QFI of the atoms is given by $%
F_{A}=N$ for the coherent spin state $\hat{\rho}_{A}=|j,-j\rangle \langle
j,-j|$, as mentioned above. When the coupling $\lambda$ increases up to its
critical point $\lambda _{\mathrm{cr}}$, a large number of bosons appears~\cite%
{EB03,Esslinger} and $F_{B}$ begins to increase. It surpasses the classical
limit around $\lambda_{\mathrm{cr}}$ as the ratio $F_{B}/(4\bar{n})>1$ [see the solid
lines of Fig.~\ref{FIG1}(a)]. From Fig.~\ref{FIG1}(b), one can note that the
QFI of $\hat{\rho}_{A}$ with small $N$ cannot beat the classical limit; The
ratio $F_{A}/N$ is always smaller than $1$ and decreases monotonically with
increasing $\lambda$. Only for large enough number of atoms (say, $N>10$), the scaled QFI $F_{A}/N$ can be larger than $1$ at $\lambda\sim\lambda_{\mathrm{cr}}$.

It is interesting to observe two key features of the finite-$N$ Dicke model:
(i) near the critical point $\lambda _{\mathrm{cr}}$, both $\hat{\rho}_{A}$ and $\hat{%
\rho}_{B}$ provide enhanced QFIs beyond the classical limits, although
they are in general highly mixed states; (ii) for $\lambda\gg\lambda _{\mathrm{cr}}$,
the QFI of the field approaches the classical limit, i.e., $F_{B}\rightarrow
4\bar{n}$, while for the atoms, $F_{A}\rightarrow 0$. To understand these
behaviors, we study in detail the quantum nature of $\hat{\rho}_{A,B}$ (see
below). In Sec. III, we further present analytical results of the QFIs in
the thermodynamic limit (i.e., $N=\infty$), and find that both $F_{B}/(4\bar{%
n})$ and $F_{A}/N$ show critical behaviors at $\lambda=\lambda_{\mathrm{cr}}$.

\begin{figure}[hpbt]
\centerline{
\includegraphics[width=0.9\columnwidth,angle=0]{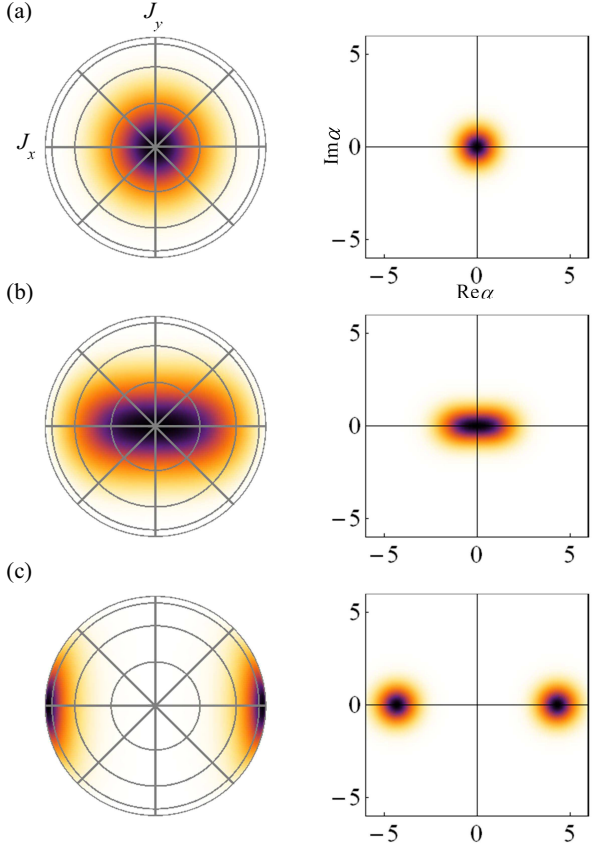}}
\caption{(Color online) Quasi-probability distributions $Q_{A}(\theta,\phi)$ (left panel) and $Q_{B}(\alpha)$ (right
panel) of the ground state for the finite-$N$ Dicke Hamiltonian with $N=20$
and the atom-field coupling strength $\protect\lambda =0$ (a), $0.54$ (b),
and $1$ (c). The axes in the Bloch sphere (left panel), i.e., the
three-dimensional phase space, $J_{x,y,z}=\langle \hat{J}_{x,y,z}\rangle$,
while for that of the field mode (right panel), $\mathrm{Re}\alpha=\langle \hat{X}_{0}\rangle$ and $\mathrm{Im}\alpha=\langle\hat{X}_{\pi /2}\rangle$, where the expectation values are taken with
respect to the coherent states $|\theta,\phi\rangle$ and $|\alpha \rangle$, respectively. Other parameters: the critical
coupling $\lambda _{\mathrm{cr}}=1/2$, the same as in Fig.~\ref{FIG1},
and the density of $Q_{A}$ is normalized by its maximal value $Q_{A,\max}$~\cite{Kitagawa,Jin09}, with $Q_{A,\max }=1$ (a), $0.557$ (b), and $0.5$ (c).}
\label{FIG2}
\end{figure}

The quantum nature of $\hat{\rho}_{A,B}$ can be visualized by the
quasi-probability distribution of the atoms $Q_{A}(\theta, \phi)=\langle\theta, \phi| \hat{\rho}_{A}|\theta, \phi\rangle$ and that of
the bosonic field $Q_{B}(\alpha)=\langle\alpha|\hat{\rho}_{B}|\alpha\rangle$, where $|\alpha\rangle=e^{\alpha\hat{b}^{\dagger}-\alpha^{\ast}\hat{b}%
}|0\rangle$ and $|\theta, \phi\rangle=e^{\eta\hat{J}_{+}-\eta^{\ast}\hat{J}%
_{-}}|j,-j\rangle$ (with $\eta=\theta e^{-i\phi}/2$) denote coherent states
of the two subsystems. Overall, there is a one-to-one correspondence between
$Q_{A}$ and $Q_{B}$, as depicted by Fig.~\ref{FIG2}. For vanishing $\lambda$, both $\hat{\rho}_{A}$ and $\hat{\rho}_{B}$ are also
minimum-uncertainty states, which exhibit isotropic quasi-probability
distributions as $Q_{A}(\theta, \phi)=\cos^{4j}(\theta/2)$ and $Q_{B}(\alpha)=\exp(-|\alpha|^{2})$ [see Fig.~\ref{FIG2}(a)]. When $\lambda$
crosses $\lambda_{\mathrm{cr}}$, from Fig.~\ref{FIG2}(b), we find that both $Q_{A}$
and $Q_{B} $ become elliptical, implying the appearance of phase-squeezed
states for the two subsystems.

\begin{figure}[tbph]
\centerline{
\includegraphics[width=0.95\columnwidth,angle=0]{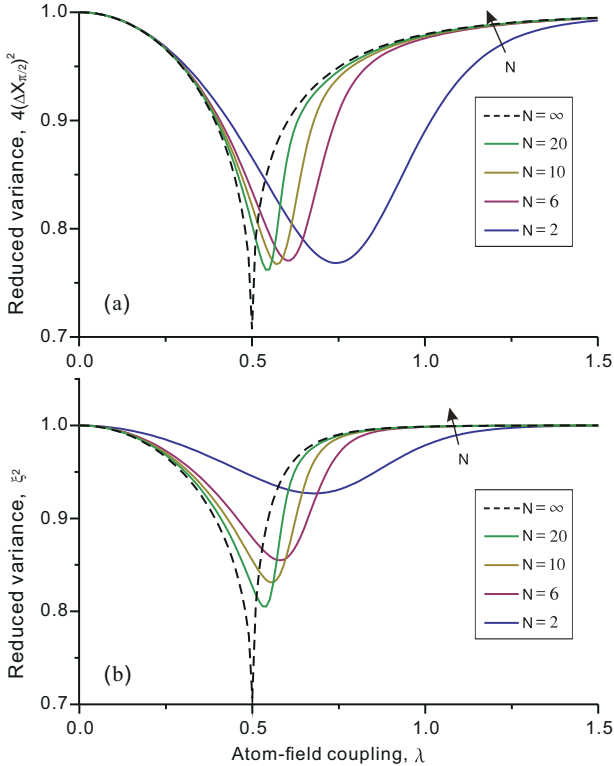}}
\caption{(color online) Degree of quadrature squeezing for the field mode $4(\Delta \hat{X}_{\pi /2})^{2}$ (a), and that of spin squeezing for
the atoms $\xi ^{2}$ (b) against the coupling strength $\lambda$ for the number of atoms $N=2$, $6$, $10$, and $20$, as indicated by
the arrow. Dashed lines: analytical results in the thermodynamic limit
(i.e., $N=\infty$), minimized at the critical point $\lambda %
_{\mathrm{cr}}=0.5$ (on resonance, as Fig.~\ref{FIG1}), which indicates that
the field and the atomic states $\hat{\rho}_{B,A}$ are nonclassical
phase-squeezed states near the critical point.}
\label{FIG3}
\end{figure}

To confirm the presence of nonclassical states at $\lambda\sim\lambda_{\mathrm{cr}}$,
we consider the quadrature squeezing of the field state $\hat{\rho}_{B}$,
following the original calculations by Emary and Brandes~\cite{EB03}. As
usual in quantum optics, we introduce a quadrature operator
\begin{equation}
\hat{X}_{\sigma }=\frac{1}{2}\left( \hat{b}e^{-i\sigma }+\hat{b}^{\dagger
}e^{i\sigma }\right) ,  \label{quadrature}
\end{equation}
where the squeezing angle $\sigma\in\lbrack 0,\pi /2]$ is to be determined.
When $\sigma=0$ or $\pi/2$, the quadrature operator represents the amplitude
or the phase component of the field mode, i.e., $\hat{X}_{0}=(\hat{b}+\hat{b}%
^{\dagger})/2$ or $\hat{X}_{\pi/2}=(\hat{b}-\hat{b}^{\dagger})/(2i)$. For
the vanishing coupling $\lambda$, the field is in the vacuum $|0\rangle$ and
hence the variance $(\Delta \hat{X}_{\sigma})^{2}=1/4$, which is the
classical limit of the field variance and is independent of the squeezing
angle $\sigma$. This isotropic variance has been depicted in the right panel of
Fig.~\ref{FIG2}(a). As the coupling $\lambda$ increases, one finds
\begin{equation*}
(\Delta \hat{X}_{\sigma })^{2}=\frac{1}{4}+\frac{\mathrm{Re}\langle \hat{b}
^{2}\rangle \cos (2\sigma )+\langle \hat{b}^{\dagger }\hat{b}\rangle }{2},
\end{equation*}
where we have used $\langle\hat{b}\rangle=0$ and $\langle \hat{b}^{2}\rangle\in\mathbb{R}$, due to the parity symmetry $\hat{\Pi}^{\dagger }\hat{b}\hat{\Pi}=-\hat{b}$ and the real atom-field coupling $\lambda$.
Minimizing $(\Delta \hat{X}_{\sigma})^{2}$ with respect to $\sigma$, we
obtain the optimal squeezing angle $\sigma_{\mathrm{op}}=0$ or $\pi/2$. Our
numerical result in Fig.~\ref{FIG2}(b) suggests $\sigma_{\mathrm{op}}=\pi/2$, which means that the optimal squeezing occurs along the $\hat{X}_{\pi /2}$ axis with the reduced variance $(\Delta \hat{X}_{\pi/2})^{2}$ smaller than
the classical limit $1/4$. In Fig.~\ref{FIG3}(a), we confirm that the degree
of squeezing $4(\Delta \hat{X}_{\pi /2})<1$ and that it is minimized at $\lambda\sim\lambda _{\mathrm{cr}}$ for large enough $N$.

Similarly, one can consider the spin squeezing of the atomic state $\hat{\rho%
}_{A}$. Due to the conserved parity, the atoms have vanishing coherence $%
\langle \hat{J}_{+}\rangle =0$ and hence the total spin $\langle \mathbf{%
\hat{J}}\rangle =(0, 0, \langle\hat{J}_{z}\rangle)$, similar to that of the
Lipkin-Meshkov-Glick model~\cite{JVidal,Tsomokos,Shi,Ma&Wang}. To quantify the degree
of spin squeezing~\cite{Kitagawa,Wineland,Jin09,Wang10,Ma,Wang12,Jia}, one can introduce a
spin component $\hat{J}_{\phi}=\hat{J}_{x}\cos\phi+\hat{J}_{y}\sin\phi$,
which is normal to the total spin. Again, the squeezing angle $\phi$ is to
be determined. Since $\langle\hat{J}_{\phi}\rangle =0$, we obtain the
variance of $\hat{J}_{\phi}$ as
\begin{equation*}
(\Delta J_{\phi })^{2}=\frac{1}{2}\left[ \langle \hat{J}_{x}^{2}+\hat{J}
_{y}^{2}\rangle +\mathrm{Re}\langle \hat{J}_{+}^{2}\rangle \cos \left( 2\phi
\right) \right],
\end{equation*}
where we have used \textrm{Im}$\langle \hat{J}_{+}^{2}\rangle\equiv\langle%
\hat{J}_{x}\hat{J}_{y}+\hat{J}_{y}\hat{J}_{x}\rangle=0$ due to the real $%
\lambda$. It is easy to find the optimal squeezing angle $\phi_{\mathrm{op}%
}=0$ or $\pi/2$~\cite{Jin09}, whereas the left panel of Fig.~\ref{FIG2}(b)
suggests $\phi_{\mathrm{op}}=\pi/2$, corresponding to the spin squeezing and
the anti-squeezing in the $\hat{J}_{y}$ and the $\hat{J}_{x}$ axes,
respectively. A spin squeezed state is defined if the reduced variance of $%
\hat{J}_{y}$ is smaller than the classical limit $N/4$~\cite{Kitagawa,Wineland,Jin09,Wang10,Ma,Wang12,Jia}. As shown in Fig.~\ref{FIG3}(b), one can see that the squeezing parameter $\xi^{2}=4(\Delta\hat{J}_{y})^{2}/N\leq 1$ and is
minimized around the critical point $\lambda_{\mathrm{cr}}$.

From the solid lines of Fig.~\ref{FIG1}, we also note that as the coupling $\lambda \gg \lambda _{\mathrm{cr}}$, the QFI of the field $F_{B}\rightarrow 4\bar{n}$, while for the atoms $F_{A}\rightarrow 0$. This behavior can be understood by examining the ground state of the finite-$N$ Dicke Hamiltonian with $\lambda \rightarrow \infty$~\cite{ultra1,ultra2,ultra3}. In this
ultra-strong coupling regime, the number of bosons $\bar{n}\varpropto
\lambda ^{2}\rightarrow \infty $ and hence the dominant term of the Dicke
Hamiltonian is given by $\hat{H}_{0}=\omega \hat{b}^{\dagger}\hat{b}%
+2\lambda (\hat{b}+\hat{b}^{\dagger })\hat{J}_{x}/\sqrt{N}$. Minimizing the
energy functional $_{x}\langle j,m|\langle \alpha |\hat{H}_{0}|\alpha
\rangle |j,m\rangle _{x}$ with respect to $\alpha $ and $m$, one can obtain
the atomic state $\hat{\rho}_{A}=(|j,+j\rangle _{xx}\langle
j,+j|+|j,-j\rangle _{xx}\langle j,-j|)/2$, where $|j,\pm j\rangle _{x}$,
being eigenvectors of $\hat{J}_{x}$, provide the variances $(\Delta \hat{J}_{x})_{\pm}^{2}=0$. Therefore, from Eq.~(\ref{QFI}) we have $F_{A}\rightarrow 0$ as $\lambda \rightarrow \infty $. Similarly, the field
state is given by $\hat{\rho}_{B}=(|+\alpha_{0}\rangle \langle +\alpha
_{0}|+|-\alpha _{0}\rangle \langle -\alpha_{0}|)/2$, where $|\pm \alpha
_{0}\rangle $, with the amplitude $\alpha_{0}=\lambda\sqrt{N}/\omega$,
denote the coherent states of bosons~\cite{ultra1,ultra2,ultra3}. It is easy
to find that as the amplitude $\alpha _{0}$ ($\varpropto \lambda $) $\rightarrow \infty $, the two coherent states $|\pm \alpha _{0}\rangle $,
almost orthogonal with each other, provide the variances $(\Delta \hat{b}^{\dagger }\hat{b})_{\pm }^{2}=\alpha _{0}^{2}=\bar{n}$. As a result, we
obtain the total QFI of the bosonic field $F_{B}\approx 4\sum_{\pm }p_{\pm
}(\Delta \hat{b}^{\dagger}\hat{b})_{\pm}^{2}=4\bar{n}$ (due to $p_{\pm
}=1/2$), leading to the ratio $F_{B}/(4\bar{n})\rightarrow 1$ as $\lambda
\rightarrow \infty $.

When the coupling $\lambda \gg \lambda _{\mathrm{cr}}$, the quasi-probability
distribution $Q_{A,B}=(I_{-}+I_{+})/2$ for each subsystem $\hat{\rho}_{A,B}$
contains two contributions, i.e., $I_{\pm}=[1\pm \sin (\theta )\cos(\phi
)]^{N}/2^{N}$ for the atoms and $I_{\pm}=\exp(-|\alpha \pm \alpha
_{0}|^{2})$ for the bosonic field, as illustrated in Fig.~\ref{FIG2}(c). In
addition, we obtain the reduced variances $(\Delta \hat{J}_{y})^{2}=N/4$
(i.e., $\xi^{2}=1$) and $(\Delta \hat{X}_{\pi/2})^{2}=1/4$ (see the solid
curves in Fig.~\ref{FIG3}), as well as the increased variances $(\Delta \hat{%
X} _{0})^{2}=\alpha _{0}^{2}+1/4$ and $(\Delta \hat{J}_{x})^{2}=N^{2}/4$,
which have been confirmed by Ref.~\cite{Hirsch}.


\section{Quantum Fisher information in the thermodynamic limit}

In this section, we first briefly review the quantum critical behavior of
the Dicke model in the thermodynamic limit based on the solution outlined by
Emary and Brandes~\cite{EB03}, and then present analytical results of the
QFIs for the field and the atomic subsystems.

The standard procedure for the diagonalization of the Dicke Hamiltonian
consists of four steps~\cite{EB03}. First, performing the Holstein-Primakoff
transformation $\hat{J}_{+}=(\hat{J}_{-})^{\dag }=\hat{a}^{\dagger }\sqrt{N-
\hat{a}^{\dagger }\hat{a}}$ and $\hat{J}_{z}=\hat{a}^{\dagger }\hat{a}-N/2$,
one can write down the Dicke Hamiltonian and the parity in terms of bosonic
operators $\hat{a}$ and $\hat{b}$. Second, the operators $\hat{a}$ and $\hat{
b}$ are decomposed as
\begin{equation}
\hat{a}=\delta \hat{a}\pm \alpha _{s}, \hskip 15pt \hat{b}=\delta \hat{b}\mp
\beta _{s},  \label{displacements}
\end{equation}
where $\alpha_{s}$ ($\beta_{s}$) denotes the mean-field part and $\delta%
\hat{a}$ ($\delta \hat{b}$) the quantum fluctuation for each subsystem.
Third, one obtains the Dicke Hamiltonian up to quadratic order in $\delta\hat{%
a}$ and $\delta\hat{b}$ by eliminating their linear terms, which gives the
solution for the mean field parts~\cite{EB03}:
\begin{equation}
\alpha _{s}=\sqrt{\frac{N(1-\mu )}{2}}, \hskip 15pt \beta _{s}=\frac{
\lambda }{\omega }\sqrt{N(1-\mu ^{2})},  \label{order parameters}
\end{equation}
where the order parameters $\alpha_{s}$ and $\beta_{s}$ are vanishing for $\mu=1$ in the
normal phase (i.e., $\lambda\leq\lambda_{\mathrm{cr}}$), and $\alpha _{s}$, $\beta
_{s}\sim O(\sqrt{N})$ for $\mu=(\lambda_{\mathrm{cr}}/\lambda)^{2}<1$ in the
superradiant phase, a unified description to the both phases~\cite{mu}.
Finally, one can diagonalize the effective Hamiltonian by introducing the
Bogoliubov transformation: $\hat{q}_{1}\!=\hat{X}_{B}\cos\gamma-\hat{X}
_{A}\sin\gamma$ and $\hat{q}_{2}\!=\hat{X}_{B}\sin\gamma+\hat{X}%
_{A}\cos\gamma$, where $\hat{q}_{1,2}$ denote the position operators of the
polaritons and $\hat{X}_{A,B}$ are that of the atoms (A) and the bosonic
field (B), i.e., $\hat{X}_{A}\propto(\delta \hat{a}^{\dagger}+\delta\hat{a})$ and $\hat{X}_{B}\propto(\delta \hat{b}^{\dagger}+\delta \hat{b})$. For the mixing angle $\gamma$ given by $\tan
(2\gamma)=4\lambda\sqrt{\omega_{0}\omega\mu}/[(\omega_{0}/\mu)^{2}-%
\omega^{2}]$, one can obtain the Hamiltonian of the two-mode polaritons, with
the excitation energies $\varepsilon_{k}$ (for $k=1$, $2$) determined by~\cite{EB03},
\begin{eqnarray}
\varepsilon _{k}^{2} &=&\frac{\omega ^{2}+\left( \omega _{0}/\mu \right)
^{2} }{2}+\frac{(-1)^{k}}{2}  \notag \\
&&\times \sqrt{\left[ \omega ^{2}-\left( \omega _{0}/\mu \right) ^{2}\right]
^{2}+(4\lambda )^{2}\omega \omega _{0}\mu },  \label{excitations}
\end{eqnarray}
where $\mu$ takes different values in the two phases~\cite{mu}. In position
space of the polaritons, the ground-state wave function is given by $\Psi
_{g}(q_{1},q_{2})\equiv\langle
q_{1},q_{2}|g\rangle=(\varepsilon_{1}\varepsilon_{2}/\pi^{2})^{1/4}\exp[-(%
\varepsilon_{1}q_{1}^{2}+\varepsilon _{2}q_{2}^{2})/2]$, which, according to
the Bogoliubov transformation, can also be expressed in the coordinates of
the atomic and the field operators $\hat{X}_{A,B}$.

The two possible shifts in Eq.~(\ref{displacements}) correspond to opposite
spatial displacements of the ground state in position space, due to $%
\hat{X}_{A}=\sqrt{\omega _{0}/\widetilde{\omega }}(\hat{x}_{A}\mp \sqrt{%
2/\omega _{0}}\alpha _{s})$ and $\hat{X}_{B}=\hat{x}_{B}\pm \sqrt{2/\omega }%
\beta _{s}$, where $\widetilde{\omega}=\omega_{0}(1+\mu)/2\mu$~\cite{mu} and $\hat{x}_{A,B}$ are the position operators before the displacements. We first consider the atomic state $\hat{\rho}_{A}$ under one choice of the
displacements. The reduced density matrix of the atoms can be obtained by integrating $\Psi _{g}\Psi _{g}^{\ast }$ over the coordinate of the field operator $\hat{X}_{B}$, as done in Ref.~\cite{Lambert}. The result has the same form as that
of a thermal oscillator (see Appendix, also Ref.~\cite{Feymann}), with unit
mass and the effective oscillation frequency
\begin{equation}
\Omega =\frac{\varepsilon _{1}\varepsilon _{2}}{\varepsilon
_{1}c^{2}+\varepsilon _{2}s^{2}}\sqrt{1+\frac{(\varepsilon _{1}-\varepsilon
_{2})^{2}c^{2}s^{2}}{\varepsilon _{1}\varepsilon _{2}}},  \label{Omega}
\end{equation}%
where we have set $c=\cos \gamma $, $s=\sin \gamma $, and $\cosh (\beta
\Omega )=1+2\varepsilon _{1}\varepsilon _{2}[(\varepsilon _{1}-\varepsilon
_{2})^{2}c^{2}s^{2}]^{-1}$, with $\beta =(k_{B}T)^{-1}$ and the Boltzmann
constant $k_{B}$. In the Fock basis of the thermal oscillator \cite{Feymann},
the reduced density matrix of the atoms can be expressed as
\begin{equation}
\hat{\rho}_{A}=\frac{e^{-\beta \hat{H}_{A}}}{\mathrm{Tr}e^{-\beta \hat{H}%
_{A}}}=\sum_{n}p_{n}|\psi _{n}\rangle \langle \psi _{n}|,  \label{rhoA}
\end{equation}%
where $\hat{H}_{A}=(\hat{P}_{A}^{2}+\Omega ^{2}\hat{X}_{A}^{2})/2$ is the
effective Hamiltonian of the thermal oscillator~\cite{Lambert}, with the
eigenvectors $|\psi _{n}\rangle \equiv (\hat{a}_{\Omega }^{\dagger
})^{n}|0\rangle /\sqrt{n!}$ and the eigenvalues $p_{n}\equiv \langle \psi
_{n}|\hat{\rho}_{A}|\psi _{n}\rangle $. Here, the momentum operator is given by $\hat{P}_{A}=i\sqrt{\widetilde{\omega }/2}(\delta \hat{a}%
^{\dagger }-\delta \hat{a})=i\sqrt{\Omega /2}(\hat{a}_{\Omega }^{\dag }-\hat{%
a}_{\Omega })$ for the annihilation operator of the atomic fluctuation $\delta
\hat{a}$ and that of the thermal oscillator $\hat{a}_{\Omega}$. The position
operator of the atoms $\hat{X}_{A}$ also has a simple relationship with that of the
thermal oscillator, $\hat{X}_{A}=(\delta\hat{a}^{\dagger}+\delta \hat{a})/%
\sqrt{2\widetilde{\omega }}=(\hat{a}_{\Omega}^{\dag}+\hat{a}_{\Omega })/%
\sqrt{2\Omega}$.

According to Ref.~\cite{EB03}, the two displacements result in doubly
degenerate and orthogonal ground states in the symmetry-broken (i.e.,
superradiant) phase, which in turn gives the two atomic states $\hat{\rho}%
_{A}^{\pm }$ for each displacement. Obviously, we can diagonalize them in
the two ortho-normalized Fock basis $\{|\psi_{n}^{\pm}\rangle\}$, with $\langle \psi_{n}^{\pm}|\psi_{n^{\prime}}^{\pm}\rangle=\delta
_{n,n^{\prime }}$ and $\langle \psi _{n}^{\pm}|\psi _{n^{\prime}}^{\mp
}\rangle =0$. The total atomic state is supposed to be an incoherent
superposition of $\hat{\rho}_{A}^{\pm }$, i.e., $\hat{\rho}_{A}=(\hat{\rho}%
_{A}^{+}+\hat{\rho}_{A}^{-})/2$. Note that the collective spin operators
under the two displacements take the form
\begin{subequations}
\label{JxJy}
\begin{align}
\hat{J}_{x}& =\frac{N-2\alpha _{s}^{2}}{2\sqrt{N-\alpha _{s}^{2}}}(\delta
\hat{a}^{\dagger }+\delta \hat{a})\pm \alpha _{s}\sqrt{N-\alpha _{s}^{2}}%
+O(N^{0}),  \label{Jx} \\
\hat{J}_{y}& =\frac{\sqrt{N-\alpha _{s}^{2}}}{2i}(\delta \hat{a}^{\dagger
}-\delta \hat{a})+O(N^{0}),  \label{Jy}
\end{align}%
and $\hat{J}_{z}=(\delta \hat{a}^{\dagger }\pm \alpha _{s})(\delta \hat{a}%
\pm \alpha _{s})-N/2$, where the terms $\sim O(N^{0})$ are neglectable in
the thermodynamic limit. Using the self-consistent condition $\langle \delta
\hat{a}\rangle =0$~\cite{EB03}, it is easy to obtain the expectation values $\langle\hat{J}_{x}\rangle_{\pm}=\pm\alpha_{s}\sqrt{N-\alpha_{s}^{2}}$
and $\langle\hat{J}_{y}\rangle_{\pm}=0$ for each $\hat{\rho}_{A}^{\pm}$.
A 50:50 weighted average over $\langle\hat{J}_{x}\rangle_{\pm}$ gives $\langle \hat{J}_{x}\rangle =\langle\hat{J}_{y}\rangle =0$ for
the total density matrix $\hat{\rho}_{A}$. As in the previous finite-$N$ case, the
squeezing parameter is given by $\xi ^{2}\!=4\langle \hat{J}_{y}^{2}\rangle
/N$, with its explicit form
\end{subequations}
\begin{eqnarray}
\xi ^{2}\! &=&\!\frac{\mu \Omega }{\omega _{0}}\frac{e^{\beta \Omega }+1}{%
e^{\beta \Omega }-1}  \notag \\
&=&\frac{\mu }{2\omega _{0}}\left( \varepsilon _{1}+\varepsilon _{2}+\frac{%
\omega _{0}^{2}/\mu ^{2}-\omega ^{2}}{\varepsilon _{1}+\varepsilon _{2}}%
\right) ,  \label{xi2}
\end{eqnarray}%
where, in the the last step, we have dropped the intermediate quantities $%
\Omega $ and $e^{\beta \Omega }$ (see Appendix). The reduced variance $%
\langle \hat{J}_{y}^{2}\rangle \propto \langle (\delta \hat{a}^{\dagger
}-\delta \hat{a})^{2}\rangle$ can also be obtained as previous work~\cite%
{EB03}. To obtain the QFI, one has to diagonalize the reduced density matrix
as Eq.~(\ref{rhoA}), and then calculate the QFI for each $\hat{\rho}%
_{A}^{\pm}$ using Eq.~(\ref{QFI}). Since both of them are the same, we obtain the total QFI of the atoms
\begin{equation}
F_{A}=\frac{N\mu \omega _{0}}{\Omega }\frac{e^{\beta \Omega }-1}{e^{\beta
\Omega }+1},  \label{QFI-atoms}
\end{equation}%
which shows an exact relationship with the reduced variance, $F_{A}\xi
^{2}=N\mu^{2}$~\cite{mu}. This finding can be used to verify that the
enhanced QFI beyond the classical limit is induced by the squeezing, i.e., $%
\xi ^{2}=\omega_{0}(\omega^{2}+\omega_{0}^{2})^{-1/2}<1$ and hence $%
F_{A}/N=\xi^{-2}>1$ at $\lambda=\lambda_{\mathrm{cr}}$.

For the field subsystem, one can obtain the reduced density matrix $\hat{\rho%
}_{B}^{\pm }$ for each displacement, similar to Eq.~(\ref{rhoA}), but with
different oscillation frequency $\Omega |_{c\leftrightarrow s}$, i.e.,
interchanging $c$ and $s$ in Eq.~(\ref{Omega}). Again, we assume that the
total field state is a mixture of $\hat{\rho}_{B}^{\pm}$, which can be
diagonalized in the Fock basis $\{|\psi_{n}^{\pm}\rangle\}$. With the
self-consistent condition $\langle\delta\hat{b}\rangle=0$, it is easy to
obtain $\langle\hat{X}_{\pi/2}\rangle =\langle\hat{P}_{B}\rangle/\sqrt{%
2\omega}=0$ \cite{EB03}, where $\hat{P}_{B}=i\sqrt{\omega/2}(\delta\hat{b}%
^{\dagger}-\delta\hat{b})=i\sqrt{\Omega /2}(\hat{b}_{\Omega}^{\dag}-\hat{%
b}_{\Omega})$ and $\hat{X}_{\pi /2}$ is the quadrature operator of the
bosonic field, defined by Eq.~(\ref{quadrature}). For each $\hat{\rho}%
_{B}^{\pm}$, we find that the reduced variances $(\Delta \hat{X}_{\pi/2})_{\pm
}^{2}=\langle\hat{P}_{B}^{2}\rangle_{\pm}/(2\omega)$ take the same form
(see Appendix), and therefore
\begin{eqnarray}
(\Delta \hat{X}_{\pi /2})^{2} &=&\frac{\Omega }{4\omega }\frac{e^{\beta
\Omega }+1}{e^{\beta \Omega }-1}  \notag \\
&=&\frac{1}{8\omega }\left( \varepsilon _{1}+\varepsilon _{2}-\frac{\omega
_{0}^{2}/\mu ^{2}-\omega ^{2}}{\varepsilon _{1}+\varepsilon _{2}}\right) .
\label{VarY0}
\end{eqnarray}%
Again, we have removed the intermediate quantities $\Omega $ and $e^{\beta
\Omega }$ in the the last step. The QFI of the field mode depends on the
matrix elements $\langle\psi_{m}^{\pm }|\hat{G}|\psi_{n}^{\pm}\rangle$,
where $\hat{G}=\hat{b}^{\dagger}\hat{b}=(\delta\hat{b}^{\dag }\mp \beta
_{s})(\delta\hat{b}\mp\beta_{s})$ denotes the phase-shift generator and $\delta \hat{b}^{\dagger}\pm\delta\hat{b}\varpropto(\hat{b}_{\Omega
}^{\dag}\pm\hat{b}_{\Omega})$, as mentioned above. After some tedious
calculations, we obtain the total QFI of the field mode (see Appendix)
\begin{equation}
F_{B}=\frac{(\omega ^{2}-\Omega ^{2})^{2}}{2\omega ^{2}\Omega ^{2}}\frac{%
(e^{\beta \Omega }+1)^{2}}{e^{2\beta \Omega }+1}+\frac{4\omega \beta _{s}^{2}%
}{\Omega }\frac{e^{\beta \Omega }-1}{e^{\beta \Omega }+1},  \label{FQI-field}
\end{equation}%
where $\beta _{s}$ is the order parameter, given by Eq.~(\ref{order
parameters}). In the normal phase, the first term of Eq.~(\ref{FQI-field})
dominates due to $\beta _{s}=0$. On the contrary, for the superradiant
phase, the first term vanishes quickly and the second term becomes important
due to the macroscopic occupation $\beta_{s}^{2}\sim O(N)\rightarrow
\infty $, which gives a simple relation $F_{B}(\Delta\hat{X}_{\pi
/2})^{2}\approx \beta_{s}^{2}\approx\bar{n}$ (see Appendix for the explicit
form of $\bar{n}$).

Our above results, Eq.~(\ref{xi2})-Eq.~(\ref{FQI-field}), dependent upon the parameter $\mu$, are valid for the both phases~\cite{mu}. In Fig.~\ref{FIG1} and Fig.~\ref{FIG3}, we plot the scaled QFIs and the reduced variances of $\hat{\rho}_{A,B}$ as a function of the atom-field coupling $\lambda $. For the vanishing coupling $\lambda $, we have $F_{A}/N=\xi^{2}=4(\Delta\hat{X}_{\pi/2})^{2}=1$ and $F_{B}=0$ (as $\Omega=\omega$), due to $\mu=1$, $\alpha_{s}=\beta_{s}=0$, and $\varepsilon_{1}+\varepsilon _{2}=\omega+\omega_{0}$. When the coupling increases up to $\lambda_{\mathrm{cr}}$, the lower-branch polaritonic energy is gapless, i.e., $\varepsilon_{1}=0$ and $\varepsilon_{2}=(\omega^{2}+\omega_{0}^{2})^{1/2}$, so the reduced variance of the field $4(\Delta \hat{X}_{\pi/2})^{2}=\omega(\omega^{2}+\omega_{0}^{2})^{-1/2}<1$ and hence $F_{B}/(4\bar{n})\approx 1/[4(\Delta \hat{X}_{\pi/2})^{2}]>1$, similar to that of the atomic subsystem. From the dashed lines of Fig.~\ref{FIG1} and Fig.~\ref{FIG3}, one can easily see that at the critical point $\lambda_{\mathrm{cr}}=1/2$ (on resonance with $\omega\omega_{0}=1$), the scaled QFIs $F_{A}/N=F_{Q}/(4\bar{n})\rightarrow \sqrt{2}$ due to $\xi^{2}=4(\Delta\hat{X}_{\pi/2})^{2}=1/\sqrt{2}$. As the coupling $\lambda\rightarrow\infty$, the excitation energies $\varepsilon _{1}\rightarrow \omega$ and $\varepsilon _{2}\rightarrow\omega_{0}/\mu$, yielding $\xi^{2}\rightarrow 1$ and $F_{A}/N\approx\mu^{2}\rightarrow 0$; While for the bosonic field, we have $\Omega\approx\omega$ and $e^{\beta\Omega}\rightarrow\infty$, so $(\Delta\hat{X}_{\pi/2})^{2}\rightarrow 1/4$ and $F_{B}/(4\bar{n})\rightarrow 1$, returning to the classical limit.

Finally, let us investigate the scaling behaviors of the QFI of the atoms $F_{A}/N$ and that of the field mode $F_{B}/(4\bar{n})$ at $\lambda \sim
\lambda_{\mathrm{cr}}$. The critical exponents of a quantum phase transition are manifested in the behavior of the excitation energies~\cite{Sachdev}. For the Dicke model, it has been shown that the lower-branch excitation energy vanishes at $\lambda=\lambda _{\mathrm{cr}}$ as $\varepsilon_{1}\sim |\lambda-\lambda_{\mathrm{cr}}|^{2\nu}$, with the critical exponent $\nu=1/4$~\cite{EB03}. Recently, Nataf~\textit{et al.}~\cite{Hur} have found that quantum fluctuations of the field $\Delta \hat{X}_{0}\Delta \hat{X}_{\pi/2}$ diverges as $|\lambda-\lambda_{\mathrm{cr}}|^{-1/4}$ near the critical point. Here, we show that the QFI of the atoms $F_{A}/N$ is nonanalytic at $\lambda =\lambda_{\mathrm{cr}}$, since its first-order derivative diverges as $\left.\partial_{\lambda }(F_{A}/N)\right\vert_{\lambda\rightarrow\lambda_{\mathrm{cr}}\pm 0}\sim\vert\lambda-\lambda_{\mathrm{cr}}\vert^{-1/2}$. A similar result can be obtained for the field mode, $F_{B}/(4\bar{n})$, indicating that the QFIs of both subsystems are sensitive to the quantum criticality of the Dicke model, as one expects.

\section{Conclusion}

In summary, we have investigated the quantum Fisher information of the
field and that of the atoms in the ground state Dicke model. For finite and
large enough $N$, we find that the QFI of each subsystem can beat the
classical limit near the critical point $\lambda_{\mathrm{cr}}$, due to the
appearance of a nonclassical squeezed state, as demonstrated
numerically by the quasi-probability distribution of $\hat{\rho}_{A,B}$ and
the reduced quadrature variance below the classical limit. When the atom-field
coupling enters the ultra-strong regime $\lambda \gg \lambda_{\mathrm{cr}}$, we find
the QFI of the bosonic field $F_{B}\rightarrow 4\bar{n}$, while for the
atoms $F_{A}\rightarrow 0$, since $\hat{\rho}_{A,B}$ at $\lambda \rightarrow
\infty$ is an incoherent mixture of two coherent states $|\pm\alpha
_{0}\rangle$ and that of the atoms $|j,\pm j\rangle _{x}$, respectively. In
the thermodynamic limit, we present analytical relations of the QFIs and the
reduced variances for both subsystems, $F_{A}\xi ^{2}=N\mu^{2}$ and $F_{B}(\Delta \hat{X}_{\pi /2})^{2}\approx \bar{n}$, which verify that the
enhanced QFI near $\lambda _{\mathrm{cr}}$ is induced by the squeezing. For each
subsystem, we find that the first-order derivative of the QFI diverges as $%
\lambda \rightarrow \lambda _{\mathrm{cr}}\pm 0$, a sensitive probe of the
superradiant quantum phase transition.

\begin{acknowledgments}
We thank Dr. J. Ma, Prof. X. Wang, Prof. Y. X. Liu, and Prof. C. P. Sun for
helpful discussions. This work is supported by the NSFC (Contract
Nos.~11174028 and 11274036), the FRFCU (Contract No.~2011JBZ013), and the
NCET (Contract No.~NCET-11-0564). FN is partially supported by the RIKEN iTHES Project, MURI Center for Dynamic Magneto-Optics, JSPS-RFBR contract No. 12-02-92100, Grant-in-Aid for Scientific Research (S), MEXT Kakenhi on Quantum Cybernetics, and the JSPS via its FIRST program.
\end{acknowledgments}

\appendix

\section{Detailed derivations of Eq.~(10)-Eq.~(13)}

In the thermodynamic limit, we calculate the reduced density matrix of the atoms by integrating $\Psi_{g}\Psi_{g}^{\ast}$ over the coordinate of the bosonic field (see e.g., Ref.~\cite{Lambert}), namely $\rho_{A}(X_{A},X_{A}^{\prime})=\int_{-\infty}^{\infty }dX_{B}\Psi_{g}(X_{B},X_{A})\Psi _{g}^{\ast }(X_{B},X_{A}^{\prime})$, which is indeed the density matrix of a single-mode harmonic oscillator at finite temperature $T$~\cite{Feymann}:
\begin{equation}
\rho _{A}\varpropto \exp \left\{ -\frac{M\Omega \left[ \cosh (\beta \Omega
)(X_{A}^{2}+X_{A}^{\prime 2})-2X_{A}^{\prime }X_{A}\right] }{2\sinh (\beta
\Omega )}\right\} ,  \label{rhoAPS}
\end{equation}%
where $\cosh(\beta\Omega)=1+2\varepsilon _{1}\varepsilon_{2}/[(\varepsilon _{1}-\varepsilon _{2})^{2}c^{2}s^{2}]$, with $c=\cos\gamma $ and $s=\sin \gamma$. By taking the mass $M=1$, we further obtain the effective oscillation frequency as Eq.~(\ref{Omega}). In the Fock basis
of the thermal oscillator, the reduced density matrix can be rewritten as Eq.~(\ref{rhoA}). For the two possible displacements, we adopt the notation $\hat{\rho}_{A}^{\pm}$ and diagonalize them as $\hat{\rho}_{A}^{\pm}=\sum_{n}p_{n}|\psi_{n}^{\pm}\rangle\langle\psi_{n}^{\pm}|$ , where $\{|\psi_{n}^{\pm}\rangle\}$ are Fock states of the thermal oscillators and $p_{n}=e^{-\beta \Omega (n+1)}/(e^{\beta \Omega}-1)$ are the weights.
It is reasonable to assume the total atomic state $\hat{\rho}_{A}=(\hat{\rho}_{A}^{+}+\hat{\rho}_{A}^{-})/2$.

We now calculate the reduced variance and the QFI of the atoms for each $\hat{\rho}_{A}^{\pm}$. Using Eq.~(\ref{rhoA}) and Eq.~(\ref{Jy}), we obtain
the expectation value $\langle\hat{J}_{y}\rangle_{\pm}=0$ for each $\hat{\rho}_{A}^{\pm}$, and also its variance
\begin{equation}
\langle \hat{J}_{y}^{2}\rangle _{\pm }=\sum_{n=0}^{+\infty }p_{n}\langle
\psi _{n}^{\pm }|\hat{J}_{y}^{2}|\psi _{n}^{\pm }\rangle =\frac{N}{4}\frac{%
\mu \Omega }{\omega _{0}}\frac{e^{\beta \Omega }+1}{e^{\beta \Omega }-1},
\label{Jy2}
\end{equation}%
where we have used $\hat{a}_{\Omega}^{\dag}\hat{a}_{\Omega}|\psi_{n}^{\pm}\rangle =n|\psi_{n}^{\pm}\rangle$, $\alpha_{s}^{2}=N(1-\mu)/2$, and the identities
\begin{equation}
\sum_{n=0}^{+\infty }np_{n}=\frac{1}{e^{\beta \Omega }-1}, \hskip 16pt
\sum_{n=0}^{+\infty }n^{2}p_{n}=\frac{e^{\beta \Omega }+1}{(e^{\beta \Omega
}-1)^{2}}.  \label{identities}
\end{equation}%
Note that the variances $\langle \hat{J}_{y}^{2}\rangle_{\pm}$ for each $\hat{\rho}_{A}^{\pm}$ are the same, so we obtain the total variance $\langle\hat{J}_{y}^{2}\rangle=\langle\hat{J}_{y}^{2}\rangle_{\pm}$ and hence the squeezing parameter $\xi^{2}=4\langle\hat{J}_{y}^{2}\rangle N$, as given by Eq.~(\ref{xi2}). The intermediate quantities $\Omega$ and $e^{\beta \Omega}$ can be removed by using the relation
\begin{equation}
\frac{e^{\beta \Omega }+1}{e^{\beta \Omega }-1}=\sqrt{\frac{\cosh \beta
\Omega +1}{\cosh \beta \Omega -1}}=\sqrt{1+\frac{(\varepsilon
_{1}-\varepsilon _{2})^{2}c^{2}s^{2}}{\varepsilon _{1}\varepsilon _{2}}},
\label{relation3}
\end{equation}%
where $\cosh\beta\Omega$ has been defined in Eq.~(\ref{rhoAPS}). Multiplying the above result with Eq.~(\ref{Omega}), we further obtain
\begin{equation}
\Omega \frac{e^{\beta \Omega }+1}{e^{\beta\Omega}-1}=\varepsilon
_{1}s^{2}+\varepsilon _{2}c^{2}=\frac{\varepsilon _{1}+\varepsilon _{2}}{2}+%
\frac{\omega _{0}^{2}/\mu ^{2}-\omega ^{2}}{2(\varepsilon_{1}+\varepsilon
_{2})},  \label{relation1}
\end{equation}%
which reduces the final result of $\xi^{2}$ as Eq.~(\ref{xi2}). According to Eq.~(\ref{QFI}), the QFI depends upon the matrix elements of the phase-shift generator $\hat{G}$ ($=\hat{J}_{x}$). For each $\hat{\rho}_{A}^{\pm}$, from Eq.~(\ref{Jx}) we obtain
\begin{eqnarray*}
\langle \psi _{m}^{\pm }|\hat{J}_{x}|\psi _{n}^{\pm }\rangle &=&\pm \alpha
_{s}\sqrt{N-\alpha _{s}^{2}}\delta _{m,n}+\frac{N-2\alpha _{s}^{2}}{2\sqrt{%
N-\alpha _{s}^{2}}} \\
&&\times \sqrt{\frac{\widetilde{\omega }}{\Omega }}\left( \sqrt{n+1}\delta
_{m,n+1}+\sqrt{n}\delta _{m,n-1}\right) ,
\end{eqnarray*}%
and thereby $\langle\psi_{n}^{\pm}|\hat{J}_{x}^{2}|\psi_{n}^{\pm}\rangle=\sum_{m}\vert\langle\psi _{m}^{\pm}|\hat{J}_{x}|\psi
_{n}^{\pm}\rangle\vert^{2}$, with its explicit form
\begin{equation*}
\langle \psi _{n}^{\pm }|\hat{J}_{x}^{2}|\psi _{n}^{\pm }\rangle =\frac{%
(N-2\alpha _{s}^{2})^{2}}{4(N-\alpha _{s}^{2})}\frac{\widetilde{\omega }}{%
\Omega }(2n+1)+\left( N-\alpha _{s}^{2}\right) \alpha _{s}^{2},
\end{equation*}%
where we have used the relation $p_{n+1}=e^{-\beta\Omega}p_{n}$ and Eq.~(\ref{identities}). Substituting the above results into Eq.~(\ref{QFI}), we
obtain the same QFI for each $\hat{\rho}_{A}^{\pm}$, which yields the total QFI of the atoms, as Eq.~(\ref{QFI-atoms}) in main text.

Next, we calculate the reduced variance and the QFI of the bosonic field. Similar with previous case, we suppose that the total field state is given
by $\hat{\rho}_{B}=(\hat{\rho}_{B}^{+}+\hat{\rho}_{B}^{-})/2$, where $\hat{\rho}_{B}^{\pm }$ for each displacement can be diagonalized in the Fock
basis $\{|\psi _{n}^{\pm }\rangle \}$, with the weights $p_{n}=e^{-\beta\Omega (n+1)}/(e^{\beta \Omega }-1)$. For each $\hat{\rho}_{B}^{\pm }$, we
obtain the same expectation value $\langle \hat{P}_{B}\rangle _{\pm }=0$, and also the variance
\begin{equation}
\langle \hat{P}_{B}^{2}\rangle _{\pm }=\sum_{n=0}^{\infty }p_{n}\langle \psi
_{n}^{\pm }|\hat{P}_{B}^{2}|\psi _{n}^{\pm }\rangle =\frac{\Omega }{2}\frac{%
e^{\beta \Omega }+1}{e^{\beta \Omega }-1},  \label{PX2}
\end{equation}%
where we have used $\hat{P}_{B}=i\sqrt{\Omega /2}(\hat{b}_{\Omega }^{\dag }-\hat{b}_{\Omega })$ and $\hat{b}_{\Omega }^{\dagger }\hat{b}_{\Omega }|\psi_{n}^{\pm }\rangle =n|\psi _{n}^{\pm }\rangle $. Therefore, the reduced variance of $\hat{\rho}_{B}$ is given by $(\Delta \hat{X}_{\pi
/2})^{2}=\langle \hat{P}_{B}^{2}\rangle _{\pm }/(2\omega )$, as the first result of Eq.~(\ref{VarY0}). The intermediate quantities $\Omega $ and $e^{\beta \Omega }$ can be removed by using the identity
\begin{equation}
\Omega \frac{e^{\beta \Omega }+1}{e^{\beta \Omega }-1}=\varepsilon
_{1}c^{2}+\varepsilon _{2}s^{2}=\frac{\varepsilon _{1}+\varepsilon _{2}}{2}-%
\frac{\omega _{0}^{2}/\mu ^{2}-\omega ^{2}}{2(\varepsilon _{1}+\varepsilon
_{2})},  \label{relation2}
\end{equation}%
where we have interchanged $c$ ($=\cos \gamma $) and $s$ ($=\sin \gamma $) in a comparison with Eq.~(\ref{relation1}). We now calculate the QFI of the bosonic field, which depends on the phase-shift generator $\hat{G}=\hat{b}^{\dagger }\hat{b}=(\delta \hat{b}^{\dag }\mp \beta _{s})(\delta \hat{b}\mp
\beta _{s})$, with the relations $\delta \hat{b}^{\dagger }-\delta \hat{b}=\sqrt{\Omega /\omega }(\hat{b}_{\Omega }^{\dag }-\hat{b}_{\Omega })$ and $
\delta \hat{b}^{\dagger }+\delta \hat{b}=\sqrt{\omega /\Omega }(\hat{b}_{\Omega }^{\dag }-\hat{b}_{\Omega })$. The matrix elements of $\hat{G}$ are
given by
\begin{eqnarray*}
\langle \psi _{m}^{\pm }|\hat{G}|\psi _{n}^{\pm }\rangle \! &=&\!\mp \beta
_{s}\sqrt{\frac{\omega }{\Omega }}\left( \sqrt{n+1}\delta _{m,n+1}+\sqrt{n}%
\delta _{m,n-1}\right)  \\
&&+\Delta _{-}\left[ \sqrt{(n+1)(n+2)}\delta _{m,n+2}+\sqrt{n(n-1)}\delta
_{m,n-2}\right]  \\
&&+\left[ (2n+1)\Delta _{+}+\beta _{s}^{2}-\frac{1}{2}\right] \delta _{m,n},
\end{eqnarray*}%
where $\Delta _{\pm }=(\omega ^{2}\pm \Omega ^{2})/(4\omega \Omega )$. Using the above result, one can easily obtain $\langle \psi _{n}^{\pm }|\hat{G}^{2}|\psi _{n}^{\pm }\rangle =\sum_{m}\vert \langle \psi _{m}^{\pm }|\hat{G}|\psi _{n}^{\pm }\rangle \vert ^{2}$, with
\begin{eqnarray*}
\langle \psi _{n}^{\pm }|\hat{G}^{2}|\psi _{n}^{\pm }\rangle  &=&\frac{%
\omega \beta _{s}^{2}}{\Omega }(2n+1)+2\Delta _{-}^{2}(n^{2}+n+1) \\
&&+\left[ (2n+1)\Delta _{+}+\beta _{s}^{2}-1/2\right] ^{2}.
\end{eqnarray*}%
Therefore, we can calculate the QFI for each $\hat{\rho}_{B}^{\pm }$. For instance, the first term of Eq. (\ref{QFI}) is given by
\begin{eqnarray*}
&&4\sum_{n}p_{n}(\langle \psi _{n}^{\pm }|\hat{G}^{2}|\psi _{n}^{\pm
}\rangle -|\langle \psi _{n}^{\pm }|\hat{G}|\psi _{n}^{\pm }\rangle \!|^{2})
\\
&=&8(\Delta _{-})^{2}\frac{e^{2\beta \Omega }+1}{(e^{\beta \Omega }-1)^{2}}+%
\frac{4\omega \beta _{s}^{2}}{\Omega }\frac{e^{\beta \Omega }+1}{e^{\beta
\Omega }-1},
\end{eqnarray*}%
and the second term
\begin{eqnarray*}
&&\sum_{m\neq n}\frac{8p_{m}p_{n}}{p_{m}+p_{n}}|\langle \psi _{m}^{\pm }|%
\hat{G}|\psi _{n}^{\pm }\rangle |^{2} \\
&=&\frac{32(\Delta _{-})^{2}}{(e^{-2\beta \Omega }+1)(e^{\beta \Omega
}-1)^{2}}+\frac{16\omega \beta _{s}^{2}/\Omega }{(e^{-\beta \Omega
}+1)(e^{\beta \Omega }-1)},
\end{eqnarray*}%
where we have used the relation $p_{n+k}=e^{-k\beta \Omega}p_{n}$ (for $k=1$, $2$) and Eq.~(\ref{identities}). To determine the scaled QFI $F_{B}/(4\bar{n})$, we also need to calculate the mean number of bosons
\begin{eqnarray}
\bar{n} &=&\sum_{n}p_{n}\langle \psi _{n}^{\pm }|\hat{G}|\psi _{n}^{\pm
}\rangle =\Delta _{+}\frac{e^{\beta \Omega }+1}{e^{\beta \Omega }-1}+\beta
_{s}^{2}-\frac{1}{2}  \notag \\
&=&\frac{1}{4\omega }\left[ \frac{s^{2}}{\varepsilon _{2}}\left( \varepsilon
_{2}-\omega \right) ^{2}+\frac{c^{2}}{\varepsilon _{1}}\left( \varepsilon
_{1}-\omega \right) ^{2}\right] +\beta _{s}^{2},  \label{nbar}
\end{eqnarray}%
where we have used Eq.~(\ref{relation2}) and%
\begin{equation*}
\frac{1}{\Omega }\frac{e^{\beta \Omega }+1}{e^{\beta \Omega }-1}=\frac{%
\varepsilon _{1}s^{2}+\varepsilon _{2}c^{2}}{\varepsilon _{1}\varepsilon _{2}%
}.
\end{equation*}%
Note that the last result of Eq.~(\ref{nbar}), also attainable by previous work~\cite{EB03}, shows clearly that the mean number of bosons $\bar{n}%
\approx \beta _{s}^{2}\sim O(N)$ in the superradiant phase since the first term $\sim O(N^{0})$ is neglectable as $N\rightarrow \infty$. Combining
Eq.~(\ref{FQI-field}) and Eq.~(\ref{nbar}), we can investigate the scaled QFI $F_{B}/(4\bar{n})$, as shown by the dashed line of Fig.~\ref{FIG1}(a),
and analyze the critical behavior of it near the critical point.


\begin{thebibliography}{99}
\bibitem{mu} Following Ref.~\cite{Hur}, we define the parameter $\mu=1$ in
the normal phase (i.e., $\lambda\leq\lambda_{\mathrm{cr}}$), and $\mu=\omega
_{0}\omega/(2\lambda)^{2}=(\lambda_{\mathrm{cr}}/\lambda)^{2}$ in the
superradiant phase (i.e., $\lambda >\lambda_{\mathrm{cr}}$), for which the
quantities, such as $\widetilde{\omega}=\omega_{0}(1+\mu)/2\mu$, the
mixing angle $\gamma$, and our results [Eq.~(\ref{xi2})-Eq.~(\ref{FQI-field}
)] keep being valid in the two phases.
\end{thebibliography}
\end{document}